\documentclass{article}
\usepackage[letterpaper,top=2cm,bottom=2cm,left=3cm,right=3cm,marginparwidth=1.75cm]{geometry}
\usepackage{graphicx}
\usepackage{amsmath}
\usepackage{graphicx}
\PassOptionsToPackage{hyphens}{url}
\usepackage[colorlinks=true, allcolors=blue]{hyperref}
\usepackage{makecell}
\usepackage{booktabs}
\usepackage{tabularx}
\usepackage{ltablex}
\usepackage{caption}
\usepackage{subcaption}
\usepackage{stackengine}
\usepackage{multirow}
\usepackage{authblk}
\usepackage[table]{xcolor}
\usepackage{abstract}
\usepackage{breakcites}

\title{Is generative AI really changing public sector work? Evidence from a multi-country survey}

\author[1,2]{Saba Esnaashari}
\author[1,3]{Tvesha Sippy}
\author[1]{Youmna Hashem}
\author[1,4\footnote{Corresponding author: \href{sukankana.chakraborty@ucl.ac.uk}{sukankana.chakraborty@ucl.ac.uk}}]{Sukankana Chakraborty}
\author[3]{Jonathan Bright}

\affil[1]{Public Policy Programme, The Alan Turing Institute, British Library, UK.}
\affil[2]{Institute for Global Prosperity, University College London, UK.}
\affil[3]{pattrn.ai, UK.}
\affil[4]{Centre for Advanced Spatial Analysis, University College London, UK.}

\date{}
\begin{document}

\maketitle


  \begin{abstract}
Artificial intelligence has the potential to transform how public services are delivered, with important implications for public service productivity. But how widely is the technology actually being used - and is it making any difference? Our survey of 3,362 public sector workers across the UK, the USA, South Africa and the EU finds widespread use of AI in every region. Our findings suggest that this spread is being driven by bottom-up individual adoption rather than top-down institutional adoption. In the EU, for example, 65\% of respondents use AI, while only 19\% say their organisation has formally adopted it. Most users are positive about its effects, with 79\% saying AI has improved their productivity. However, only 29\% say it has increased what is expected of them at work, raising questions about how truly transformative the technology is so far. Across regions, South Africa is a clear outlier, with the highest levels of use, organisational adoption, optimism and awareness of guidelines.
\end{abstract}

\section{Introduction}

The emergence of generative artificial intelligence (AI) has the potential to change the nature of public services, by enhancing productivity and reducing the time public sector workers spend on routine administrative tasks. This opportunity is critical for the contemporary public sector, which faces sustained fiscal pressure and rising demand for quality public services (\cite{dsit2025aiopportunities, wirtz2019artificial, mergel2019digital, oecd2025governing, worldbank2021aipublicsector}). However, the public sector has historically struggled to turn digital capability into productivity, held back by legacy systems, regulation and organisational inertia \cite{dunleavy2006digital, cordella2015egovernment}. In the UK, for example, public service productivity grew by just 0.2\% a year between 1997 and 2019, despite two decades of digital transformation programmes \cite{ons2025publicsectorproductivity}.  Despite, recent growth (1.4\% in 2023), productivity remained around 4\% below its pre-pandemic level \cite{ons2026publicsectorproductivity}. An important question, therefore, is whether contemporary AI technologies will actually realise the productivity gains expected of them. A growing body of empirical research has begun to measure these effects in real-world settings, looking at who uses AI tools, how far productivity rises, and what happens to skills \cite{peng2023impact, brynjolfsson2025generative, dellacqua2023navigating, noy2023experimental}. However, much of this work has happened within the private sector, which has a different set of dynamics and expectations to the public sector. 

A prior survey in 2023 found that a year after ChatGPT's release, 32\% respondents were using AI at work, and most ($>70\%$) said it made them more productive. This is despite the fact that only 32\% of respondents said their workplace gave clear guidance on its use \cite{bright2025generative}. This suggests that AI models are taken up by individual workers, at their own discretion. This kind of `bottom-up' adoption can bypass institutional barriers, and may mean that AI has different adoption dynamics to previous waves of technology. Unlike many earlier public sector technologies introduced through large-scale, top-down programmes, AI tools can be deployed directly by individual workers, enabling them to experiment with and automate elements of their own workflows \cite{bright2025generative}. This decentralised mode of adoption may lower barriers to use, accelerate diffusion, and allow productivity gains to emerge organically from frontline practices. However, bottom-up adoption brings costs as well. The freedom that lets a worker pick how to use a tool also leaves them to judge its risks alone. The time a tool saves can also be absorbed by rising expectations from senior management. Ultimately, the success of such bottom-up adoption, and its effects, is likely to depend on a range of factors, chief among them being the attitudes of those expected to use these technologies in their day-to-day work. While private sector studies show rapid individual-level uptake (e.g., \cite{bick2024rapid, microsoft_2025}), empirical evidence on these attitudes remains limited within the public sector, particularly across diverse international jurisdictions. 

Furthermore, existing literature largely focuses narrowly on isolated national contexts. As a result we lack evidence on how adoption rates vary across jurisdictions, departments, and demographic characteristics. How public servants perceive the direct impact of AI on their daily tasks and productivity is another area we lack granular insight on. Further, while governments have published ethical frameworks and risk management policies, empirical data is missing on whether frontline workers are actually aware of these guidelines, and what their overall concerns are about these technologies.

This paper seeks to address these gaps by presenting findings from a multi-country study examining public sector employees' attitudes towards the adoption and use of AI. We address three primary questions:

\begin{itemize}
    \item \textbf{RQ1 (Adoption and use):} How widely are public sector workers using AI, and generative AI in particular? How does this vary across countries, professions, roles and demographic groups, and how far has individual use run ahead of organisational adoption?
    \item \textbf{RQ2 (Impact on work):} How do workers perceive generative AI's effect on their productivity, the scope of their work and what is expected of them?
    \item \textbf{RQ3 (Concerns and governance):} What concerns do workers hold about security, privacy, bias and job security, do users and non-users differ, and how aware are workers of the AI guidance that applies to them?
\end{itemize}

To answer these questions, we surveyed 3,362 public sector workers across four regions (UK, USA, South Africa, and EU) in March 2025. These regions represent contrasting regulatory regimes and different levels of digital maturity, providing a more holistic picture of AI in the public sector. We asked about three types of AI: generative AI (such as ChatGPT), decision-support AI (such as triage systems) and perceptive AI (such as facial recognition).

We find that AI is now part of everyday work for most public sector employees in our sample: 74\% have used some form of AI at work, and 68\% have used generative AI. In the UK, generative AI use has risen from 22\% in our 2023 survey to 55\%. Yet individual use has run far ahead of institutional adoption, most starkly in the EU, where 65\% use generative AI but only 19\% say their organisation has adopted AI. Around 80\% of users say it makes them more productive, but nearly a third say it has also raised what is expected of them. More than half of all respondents say there is not enough guidance, and awareness of specific national guidelines is low.

The remainder of this paper is structured as follows. Section~2 describes the survey design, sampling strategy, and data used in the analysis. Section~3 presents the empirical findings, organised around adoption, usage, productivity impact, and governance-related concerns, drawing comparisons across the countries, sectors, and demographic groups studied. Section~4 discusses the implications of these findings for policy and practice.

\section{Methodology}
Data collection for this work took place online in March 2025 through a survey that was created and administered using Qualtrics. Participants were recruited through Prolific, had to be working in the public sector, and had to be based in one of ten countries: the United Kingdom, United States, South Africa, Spain, Portugal, Poland, Italy, Ireland, Germany, or France. Additionally, participants were required to be over the age of 18 to take part. Screening questions at the start of the survey confirmed that respondents met these criteria before proceeding. Participants were paid at a rate equivalent to approximately £1.20 for completing the survey, which took around seven minutes. Each participant was permitted to complete the survey only once. An attention check was included within the survey, and responses were not recorded for those who failed the check.

A total of 3,362 individuals completed the survey. Sample sizes varied by country, ranging from 59 respondents in France to around 1000 each in the United Kingdom and the United States. Respondents provided demographic information including age, gender, tenure in the public sector, current role, supervisory status, whether they worked in a frontline or non‑frontline capacity, and their average weekly working hours. 
64\% of the respondants were under 40, 64\% were women and 63\% were frontline workers. Among professions, 34\% were in Education, 20\% in Healthcare, 10\% in Public Security and Safety. The rest of the professions compromised less than 10\% of the respondents. 
Participation was limited to individuals currently employed in the public sector in the country for which the survey was administered.

\begin{table}[htbp]
\centering
\small
\setlength{\tabcolsep}{10pt}
\begin{tabular}{llrrrr}
\toprule
\textbf{Variable} & \textbf{Value} & \textbf{UK} & \textbf{EU} & \textbf{SA} & \textbf{USA} \\

 &  & n=999 & n=568 & n=845 & n=950 \\
 &  & (\%) & (\%)  & (\%)  & (\%)  \\

\midrule

\multirow{5}{*}{Age}
    & Under 30                          & 12 & 41 & 52 & 19 \\
    & 30--40                            & 37 & 36 & 32 & 35 \\
    & 41--50                            & 27 & 15 & 8 & 25 \\
    & 51--60                            & 18 & 7  & 6  & 16 \\
    & Above 60                          & 6  & 1  & 2  & 5 \\
\midrule

\multirow{3}{*}{Gender}
    & Female                             & 68 & 52 & 73 & 56 \\
    & Male                               & 31 & 46 & 26 & 43 \\
    & Non-binary                         & 1  & 2  & 0  & 1 \\

\midrule

\multirow{10}{*}{Profession}
    & Education                          & 29 & 41 & 29 & 39 \\
    & Healthcare and Public Health       & 25 & 16 & 25 & 13 \\
    & Public Safety and Security         & 10 & 10 & 6  & 12 \\
    & Legislation and Governance         & 8  & 9  & 5  & 9 \\
    & Technology and Digital Services   & 3  & 5  & 11 & 4 \\
    & Social Welfare and Housing        & 11 & 4  & 2  & 4 \\
    & Infrastructure and Transportation  & 3  & 3  & 6  & 6 \\
    & Economic and Financial Management  & 4  & 5  & 5  & 4 \\
    & Environment and Natural Resources  & 2  & 3  & 5  & 5 \\
    & Other                              & 5  & 5  & 6  & 3 \\
\midrule

\multirow{5}{*}{Government type}
    & Central (federal) government                 & 21 & 32 & 18 &  19    \\
    & Local government                   & 47 & 26 & 35 & 33 \\
    & State (province) government        &   &  33 &  44    & 45 \\
    & Public bodies or charities         & 22 &    &    &      \\
    & Other                              & 10 & 10 & 2  & 4 \\
\midrule

\multirow{2}{*}{Frontline worker}
    & Frontline role                     & 63 & 55 & 68 & 64 \\
    & Non-frontline role                 & 37 & 45 & 32 & 36 \\
\midrule

\multirow{2}{*}{Supervisor worker}
    & Non-supervisor                     & 55 & 74 & 49 & 65 \\
    & Supervisor                         & 45 & 26 & 51 & 35 \\

\bottomrule
\end{tabular}
\caption{Sample Characteristics by Region (in \%)}
\label{tab:sample_characteristics}
\end{table}

The questionnaire was identical across countries, with minor localisation adjustments where relevant and necessary, such as in the description of local government bodies. The survey was administered in English across all 10 countries.

We distinguished between three key types of AI systems and provided respondents with a definition of each type:

\begin{table}[ht]
    \centering
    \renewcommand{\arraystretch}{1.3}
    \begin{tabularx}{\linewidth}{@{}p{0.25\linewidth}X@{}}
        \textbf{Type} & \textbf{Description} \\[0.5em]

        Decision-support AI &
        Systems that directly support medical decision-making by,
        for example, identifying medical conditions or predicting
        risk. \\[0.5em]

        Perceptive AI &
        Systems that try to optimise the internal workings of a
        healthcare service by, for example, predicting missed
        appointments or forecasting staff levels. \\[0.5em]

        Generative AI &
        Systems such as ChatGPT that create text or images, that
        could be used, for example, to draft responses to patient
        queries. \\
    \caption{Types of AI Systems}
    \label{tab:ai-types}
    \end{tabularx}
\end{table}

\newpage
The survey comprised five sections: demographics; workload and administrative tasks (including average hours worked per week and time spent on administrative tasks); AI usage (covering awareness and use of decision‑support, perceptive, and generative AI systems); tasks for which AI is used (only shown to participants indicating use of generative AI); and attitudes toward AI (including questions tailored to those who use any form of AI, and questions asked of all participants). 

Participants were asked to indicate whether they were using any type of AI systems, or if they have used AI in the past but are no longer using it. All questions including tasks for which AI is used and attitudes of AI users are asked of participants who indicate that they have had some experience of using AI (whether actively using, or have used in the past). Additionally, we asked participants who stopped using AI why they did so; we report their reasons in the Result section. 

An attention check was embedded midway through the questionnaire to ensure data quality. Responses from participants who failed the attention check or did not complete the survey were removed prior to analysis. Any incomplete response - whether due to the participant dropping out midway through the survey, failing the attention check, not providing consent, or not meeting the eligibility criteria - was deleted.

Data analysis was conducted using the R programming language and includes descriptive statistics and regression analysis. The comparisons based on the presented statistics are supported by significant coefficients in regression analysis of responses on demographic characteristics, respondents' geographic location, and AI use. All data were stored securely on institutional servers, with access restricted to the project team. 

It is important to note some limitations within this study. The absence of translations for non‑English speaking countries may have introduced minor comprehension differences. The sample is subject to self‑selection bias, as individuals on Prolific may be more technologically engaged and thus more likely to use AI tools. The sample is not representative of the broader public sector workforce in each country, and the online‑only nature of the survey may exclude individuals with limited internet access or digital literacy. It is worth noting that the results including the country‑level comparisons are descriptive and no weighting or post‑stratification was applied due to the different sample structures across the countries included.
\section{Results}

\subsection{AI adoption rate}
\begin{figure}
    \centering
    \includegraphics[width=1\linewidth]{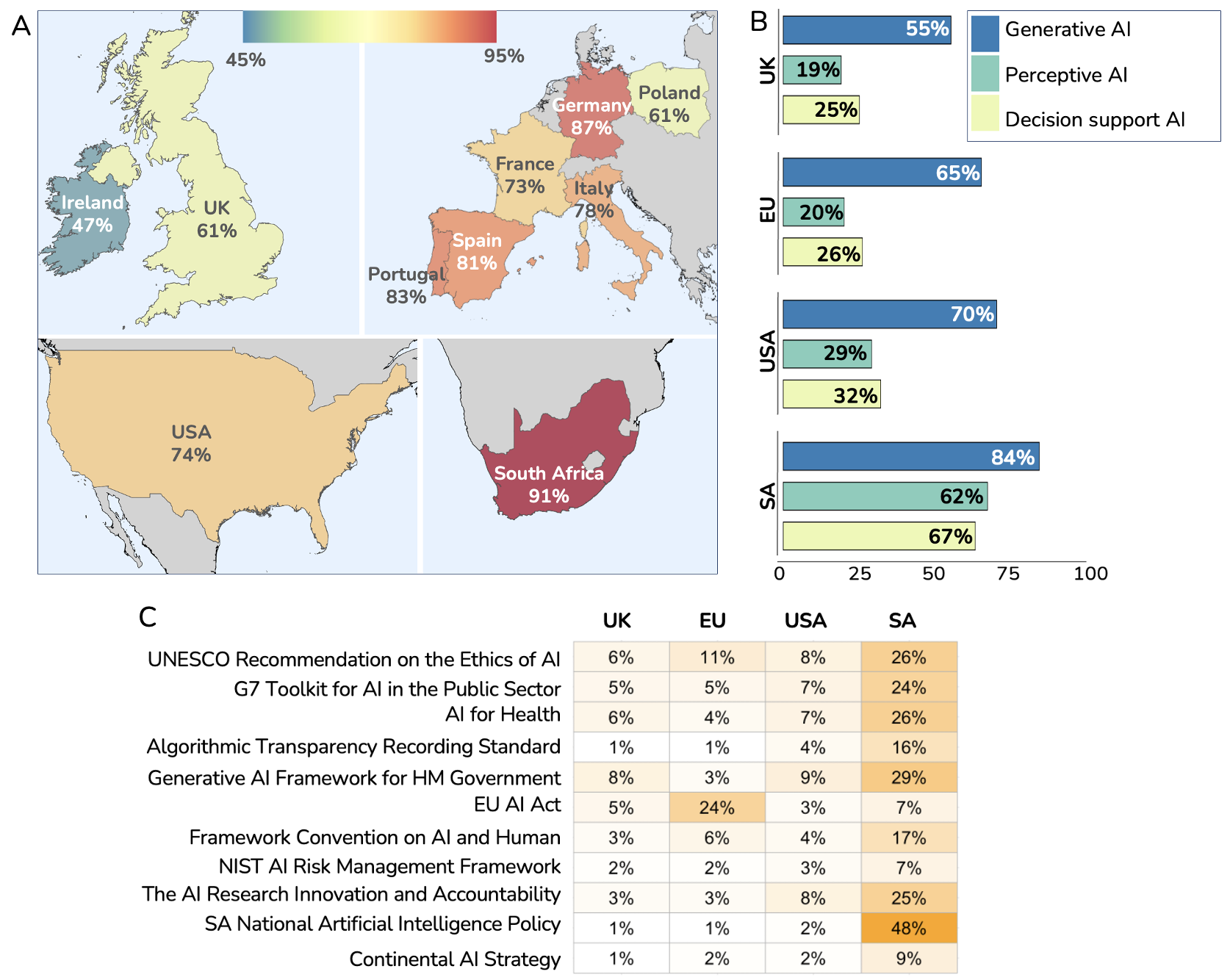}
    \caption{AI adoption by profession, Panel B: AI use by country, Panel C: Awareness of AI guidelines across regions}
    \label{fig:adoption}
\end{figure}

Figure 1 shows that AI is already a routine part of public sector work. Across our sample, 74\% of respondents have used some form of AI in their job: 63\% are current users, and a further 11\% have used it in the past but no longer do so. Adoption is high in every region we surveyed. South Africa leads, with 91\% of respondents using some form of AI, followed by the EU and the USA (both 74\%) and the UK (61\%). Among the three types of AI, we find that generative AI has significantly more users than decision-support AI and perceptive AI across all regions. GenAI adoption stands at 84\% in South Africa, 70\% in the USA, 65\% in the EU and 55\% in the UK. The other forms of AI have spread much less widely in all regions except South Africa, where adoption rates are nearly twice as much as other regions. For example, 67\% and 62\% of respondents in South Africa use perceptive AI and decision-support AI, respectively. While, in the UK, EU and USA, perceptive AI is used by between 19\% and 29\% of respondents, and decision-support AI by between 25\% and 32\%. 

Across the sample as a whole, AI use varies far more with age, role, and profession than with gender. Male respondents reported marginally higher use than female respondents (77\% vs. 73\%). Age shows a clearer variation: adoption declines from 84\% among those under 30 years to 66\% among those above 60 years. This is broadly in alignment with wider literature that finds older adults less likely to use AI tools. In the United States, for example, 66\% of 18–29-year-olds report ever using AI chatbots, compared with 23\% of those aged 65 and over \cite{pew2026age}. In the UK, public attitude surveys find that the oldest respondents report markedly lower awareness of, and experience with, AI applications \cite{modhvadia2023}.

Table \ref{tab:ai_use_characteristics} in the Appendix shows that supervisors  report higher use of AI than non-supervisors (84\% vs 68\%), and frontline workers report slightly higher use than those in non-frontline roles (77\% vs 71\%). The largest differences, however, are between professions: 93\% of respondents in Technology and Digital Services use AI, compared with 53\% of those in Social Welfare and Housing.

\subsection{Organisational adoption}

\begin{figure}
    \centering
    \includegraphics[width=1\linewidth]{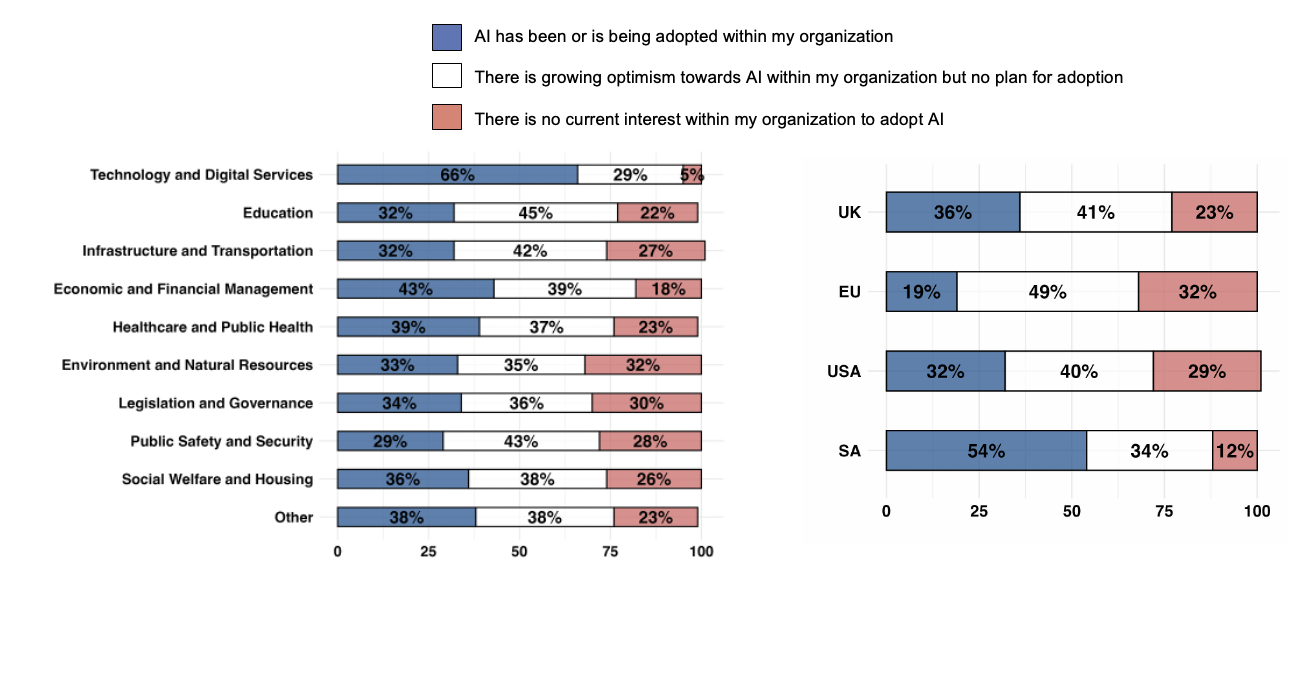}
    \caption{Organisational AI adoption. Left: Adoption by profession. Right: Adoption by country}
    \label{fig:org_adoption}
\end{figure}

To gain insights into whether adoption is bottom-up or top-down,  we asked respondents whether AI had been adopted in their organisation  -  Figure \ref{fig:org_adoption} suggests the former. In every region, individual adoption exceeds organisational adoption. Organisational adoption is at 36\% in the UK, 32\% in the USA, only 19\% in the EU, and just over 50\% in South Africa. What is striking is that, across all professions and the four regions, a sizeable portion of respondents report that there is growing optimism towards AI within their organisations, but no plan for adoption.  

\begin{figure}
    \centering
    \includegraphics[width=1.1\linewidth]{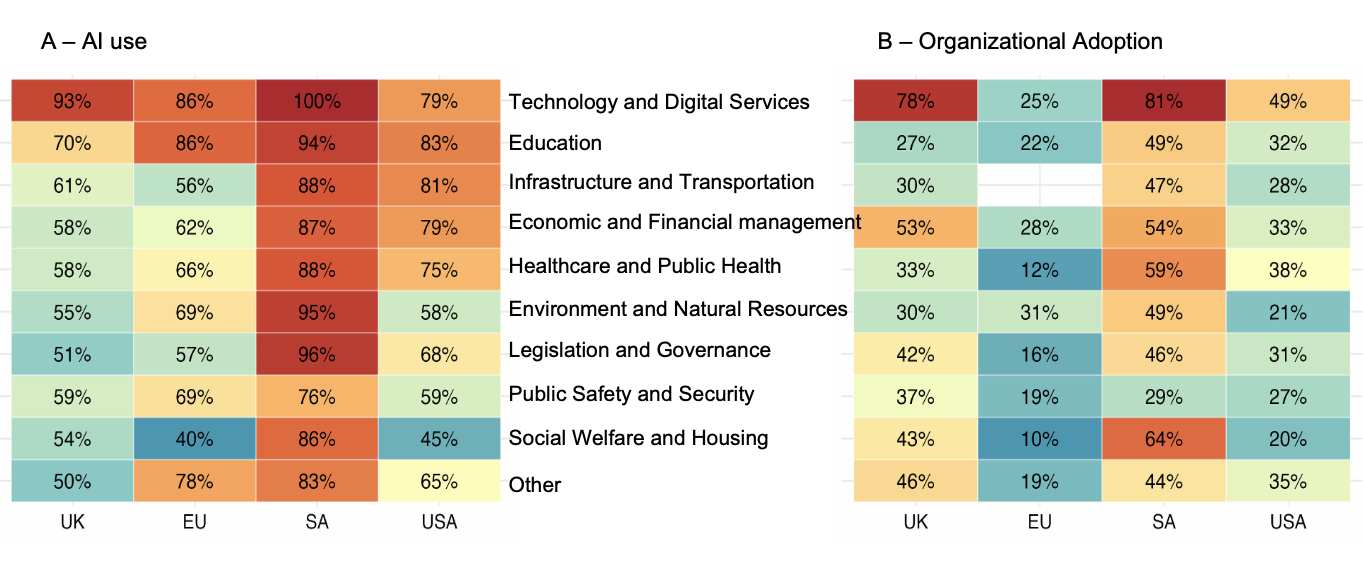}
    \caption{AI use and organisational adoption by region and profession, Panel A: AI use by profession and region, Panel B: Organisational Adoption by profession and region}
    \label{fig:org_adoption_prof_country}
\end{figure}

The same pattern appears across departments too. As might be expected, organisational adoption is highest in Technology and Digital Services, where 66\% of respondents report that AI has been adopted, and in Economic and Financial Management department (around 45\%). In every other profession it is below 40\%. Comparing these figures with rates of individual use in Table \ref{tab:ai_use_characteristics} in the Appendix shows that the professions in which workers use AI most are not necessarily those whose organisations have adopted it. Consider the case of education in the EU, which has the biggest gap between individual and organisational adoption. Figure \ref{fig:org_adoption_prof_country} shows that AI use in education across the EU region is 86\%, while organisational adoption is 22\%. Much of the adoption we observe, then, appears to be led by workers rather than by their employers. This is the bottom-up pattern our earlier study observed \cite{bright2025generative}, now observed across four regions.

\subsection{Generative AI usage}

\begin{figure}
    \centering
    \includegraphics[width=1\linewidth]{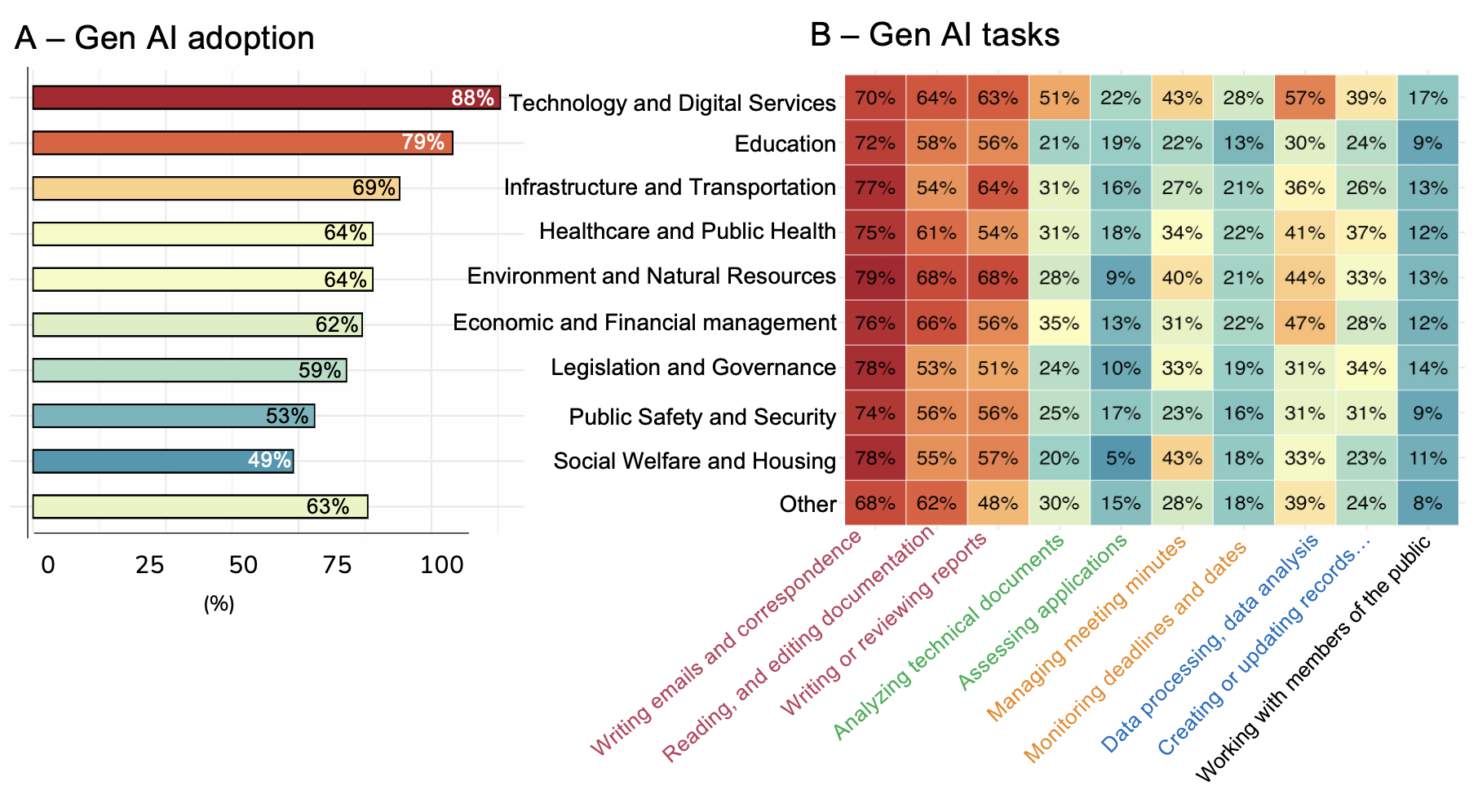}
    \caption{Panel A: Generative AI adoption by profession,  Panel B: Tasks used by profession}
    \label{fig:adoption_task_prof}
\end{figure}

We now examine patterns in individual use of Generative AI. Across regions, between 65\% and 80\% of generative AI users report using it daily or weekly. In South Africa, half of the users engage with generative AI daily and a further 39\% weekly, the highest sustained-use rate of any region. In the UK, EU, and USA, weekly use is most common (around 40\%), followed by daily use (20–35\%). When we examine use across professions, we find that those working in Technology and Education report the highest levels of generative AI use (88\% and 79\% respectively) while those in Social Welfare and Security report lowest rates, 49\% and 53\% respectively (see Figure \ref{fig:adoption_task_prof}). 

 It is also key to understand what types of tasks generative AI is being used for in the public sector. Figure \ref{fig:adoption_task_prof} shows a wide array of tasks currently being performed using generative AI. Interestingly, internal office tasks are where AI is seeing the most adoption. More than 70\% of respondents use it for "writing emails" and more than half of them use it for``reading, reviewing and editing document or paperwork", and ``writing or reviewing reports". "Working directly with members of the public" is lower down on the list. We see regional differences: employees who are based in South Africa tend to use generative AI more for most tasks compared to other regions. Across all professions too, we find high use of generative AI in administrative tasks such as reading, writing and editing documents (reports and correspondence). Many professions also use it for other administrative tasks. For example, respondents in Social Welfare and Housing in particular, despite overall low levels of generative AI use, report high levels of generative AI use for managing meeting minutes. Use of generative AI for specialised tasks such as data related tasks (processing and analysis) has highest reported use in Technology and Digital Services, Economic and Financial Management, Environmental and Natural Resources, and Healthcare and Public Health. Finally, for tasks that involve working directly with the public, we find the lowest levels of adoption across all professions. 

The demographic correlates of generative AI use broadly mirror those of AI use overall. We find that a marginally larger share of male respondents use generative AI compared to our female respondents (70\%  v 67\%). When examining use across different age groups we find that generative AI use decreases with age. Highest levels of use (77\%) are reported by respondents under the age of 30, following which usage gradually decreases with age group: 30-40 years (68\%), 41-50 years (65\%), 51-60 years (59\%) and finally 56\% for those above 60 years.  When looking at different roles, we find that those in supervisory positions use generative AI more than those in non-supervisory positions (77\% to 62\%). The difference in use between frontline and non-frontline workers was not significant. For further information, please see Table \ref{tab:ai_use_characteristics} in the Appendix. 

ChatGPT remains the most widely used tool, named by at least half of generative AI users, followed by Microsoft Copilot and Google Gemini. This marks a change from 2023, when 63\% of UK users named ChatGPT and the next most popular system, Gemini (then known as Bard), was named by only 5\% \cite{bright2025generative}. In short, generative AI is used frequently, mainly for administrative work, and increasingly across a range of tools. The next section considers the impact of this use on people's work.

\subsection{Impact of generative AI on productivity}
As outlined in the Methodology, we asked two sets of questions to understand respondents' opinions on generative AI. The first set was only shown to respondents who reported having had some experience using generative AI, and the second set was shown to all respondents, irrespective of specific type AI use. 

\begin{figure}
    \centering
    \includegraphics[width=1\linewidth]{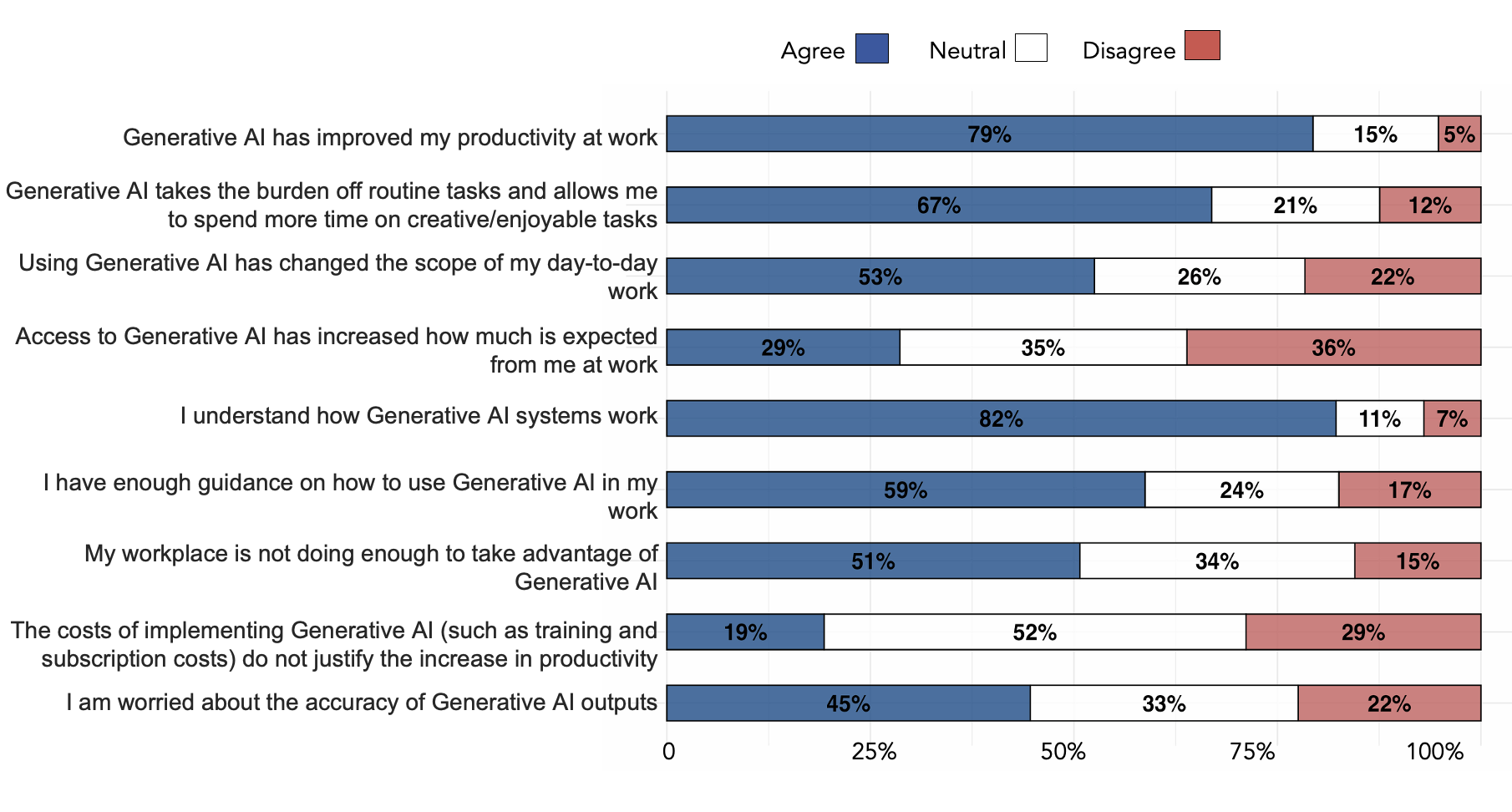}
    \caption{Opinions on AI for all AI users}
    \label{fig:ai_use_opinion}
\end{figure}

Around 80\% of the users reported a positive effect of generative AI on their productivity at work (Figure \ref{fig:ai_use_opinion}). On the other hand, around 30\% mentioned that generative AI has increased how much is expected from them at work (figure \ref{fig:ai_use_opinion}). However, this is not even in different regions as shown in Figure \ref{fig:AI_user_region_opinion} in Appendix.

South African users are generally the most positive about generative AI and its effects on their work. Yet they are also by far the most likely to report rising expectations: more than half say generative AI has increased how much is expected of them, compared with fewer than 20\% in the UK, EU and USA. South African users are also much more likely to say that generative AI has changed the scope of their day-to-day work (more than 70\%, against around 40\% in the other three regions - see Figure 9 \ref{fig:AI_user_region_opinion}). In other words, the region where generative AI is most embedded is also the one where it has done most to reshape, and to intensify, everyday work.

A similar pattern appears across professions. Figure \ref{fig:ai_opinion_prof} in the appendix shows that nearly half of users in Technology and Digital Services say generative AI has increased what is expected of them, against around 20\% in other professions.  Differences by role are smaller but still notable (Figure \ref{fig:AI_user_supervisor_opinion}). Supervisors are more likely than non-supervisors to report that generative AI has increased their productivity (85\% vs 75\%). This may reflect the kinds of task supervisors use it for, such as drafting, summarising and reporting, where gains are most visible. By contrast, men and women who use generative AI assess its effects in much the same way.

Taken together, these findings show that generative AI is widely experienced as making public sector work more productive, as it was in 2023. What has emerged since is a second, less visible effect. The same technology that saves public sector workers time has also raised what is expected of them. We return to the implications of this in Section 4. In what follows, we turn to the views of all respondents, users and non-users alike, on the risks and governance of AI.

\subsection{Perceptions and concerns about generative AI}
\begin{figure}
    \centering
    \includegraphics[width=1\linewidth]{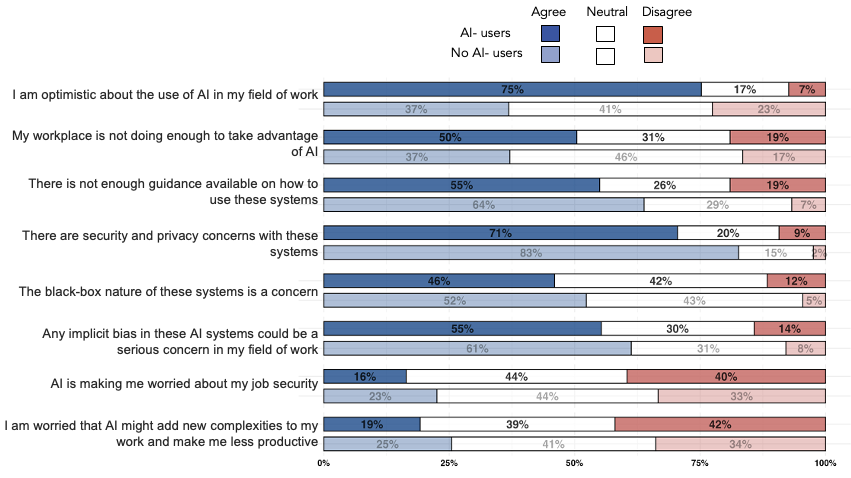}
    \caption{Opinions of all respondents divided by AI vs non-AI users}
    \label{fig:opinion_all}
\end{figure}

The second set of questions were asked from all respondents, aimed at gauging their opinion about generative AI. These questions included a vast number of topics from positive impacts to concerns about the risks of using AI on job security, as shown in Figure \ref{fig:opinion_all}. More than 70\% of respondents (both among AI users and non-AI users) showed concerns around security and privacy of AI systems and more than half of them stated that there is not enough guidance available on how to use these systems. In addition, more than half of respondents stated that any implicit bias in the AI systems could be a serious concern in their field of work. 

Perhaps unsurprisingly, we observe significant differences in the opinions of AI users vs non-AI users: 75\% of AI users were optimistic about use of AI in their field of work, while below 40\% of non AI users mentioned optimism about use of AI in work (more than 70\% of respondents were using AI in our sample). We also see slightly more concern around job security and bias issues among non-AI users. 

We observe some regional differences in the opinion of people, as shown in Figure \ref{fig:all_opinion_country} in the Appendix: more than 80\% of the respondents from South Africa are optimistic about the use of AI in work compared to between 50\%-65\% of those in other regions. 
Respondents from European countries reported more concerns about the security and privacy of these systems (81\%) compared to those from South Africa (62\%), and those in the UK reported a lack of guidance on how to use these systems as being a concern (65\%), much more than those in South Africa (47\%) or even the USA (55\%). This difference around guidance is interesting when we consider Figure \ref{fig:adoption}, where respondents from South Africa are more likely to be aware of the guidance on using generative AI. 

We also observe some differences between opinions of those who work in different professions as shown in Figure \ref{fig:ai_opinion_prof} in the Appendix: nearly 70\% of those who work in Social Welfare and Housing are concerned about any bias in AI systems compared to just 50\% of those in Technology and Digital services (generative AI adoption rate is just below 50\% in the Social Welfare profession compared to nearly 90\% in the Technology profession). 

In addition, we found differences in opinions between those who are in supervisory roles vs those who are non-supervisors as shown in Figure \ref{fig:all_opinion_supervisor} in the Appendix: for example, supervisors are more optimistic about use of AI in their field of work (73\% vs 60\%). We did not see any significant differences between males and females apart from slightly more optimism among male respondents.

\subsection{Awareness of AI guidelines}

\begin{figure}
    \centering
    \includegraphics[width=1\linewidth]{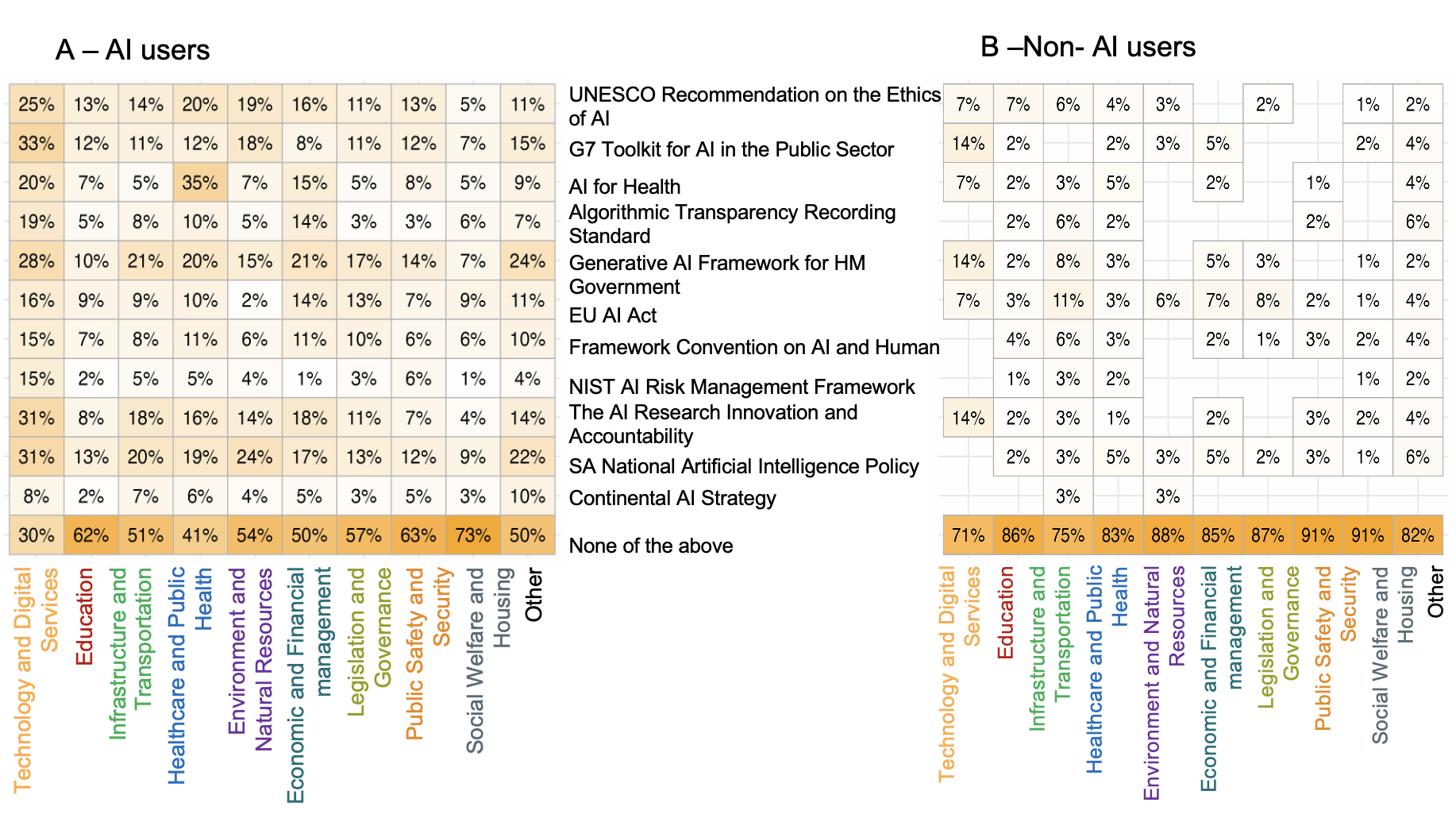}
    \caption{Guideline awareness by profession,  Panel A: AI users, Panel B: non-AI users}
    \label{fig:adoption_guideline_prof}
\end{figure}

There is growing discussion around the need for guidelines governing AI use, to ensure such tools are applied safely and without bias. This is particularly critical for public sector roles, where decisions made with the support of AI systems can directly affect citizens' access to services, benefits, and legal protections, and where errors or embedded bias carry consequences for public trust and accountability that go beyond those in most private sector contexts.

Consistent with this concern, more than half of respondents stated that there is not enough guidance available on how to use these systems (see Figure \ref{fig:opinion_all}). To understand what lies behind this perception, we test respondents' awareness of specific AI guidelines. We select two to three prominent AI guidelines from each country and region and observe awareness across all four regions. Our list contains three international guidelines.

With this in mind, we test the awareness of AI guidelines among respondents. We select between 2 to 3 prominent AI guidelines from each country and region in order to gauge awareness of these. Our list included three international guidelines: UNESCO Recommendation on the Ethics of AI, G7 Toolkit for AI in the Public Sector, and AI for Health. It also included two UK-specific guidelines: the Algorithmic Transparency Recording Standard  and the Generative AI Framework for HM Government; two EU-specific guidelines: the EU AI Act and the Framework Convention on AI and Human Rights, Democracy and the Rule of Law; two US-specific guidelines NIST AI Risk Management Framework and the AI Research Innovation and Accountability Act, and two South Africa-specific guidelines: the SA National Artificial Intelligence Policy Framework and the Continental AI Strategy.  

Figure \ref{fig:adoption} shows relatively high levels of awareness for the three international guidelines - UNESCO Recommendation on the Ethics of AI, G7 Toolkit for AI in the Public Sector, and AI for Health - across all regions. We also find that there is generally more awareness about guidelines from the respondents' own country (or region). Among UK respondents we observe relatively higher levels of awareness about the UK Generative AI Framework for HM Government (8\%). Similarly we observe 25\% of the EU respondents were aware of the EU AI Act, but there is less awareness about this guideline across other regions. Among respondents from the USA we observe high levels of awareness for the US-based AI Research Innovation and Accountability Act (8\%) and also the UK Generative AI Framework for HM Government (9\%). More strikingly, nearly 50\% of South Africa respondents were aware of the SA National Artificial Intelligence Policy Framework, but awareness of this guideline across other regions was significantly low. Overall, we also observe the highest levels of awareness across all guidelines within respondents from South Africa, and nearly 30\% of the respondents were aware of all the international guidelines, as well as the UK Generative AI Framework for HM Government and the US-based AI Research Innovation and Accountability Act. It is important to note that such self-reported awareness could be influenced by social desirability bias amongst respondents.  

Additionally, across all respondents, we find that supervisors tend to be more aware of guidelines compared to non-supervisors. Similarly, non-frontline workers report higher levels of awareness compared to frontline workers. Younger respondents (under 30) and males also report higher levels of guidelines awareness. Awareness across professions also varies (see Appendix \ref{fig:adoption_guideline_prof}). Interestingly, we find that while more than half of all respondents in each profession were not aware of any guidelines at all, this figure was only 33\% for those working in Technology and Digital Services, implying that nearly 70\% of respondents in Technology and Digital Services are aware of the given guidelines. We also find high levels of awareness for most guidelines among those working in Healthcare and Public Health, Infrastructure and Transport and Economic and Financial management.
\section{Discussion}

Our findings show that generative AI is now widely embedded in the everyday work of public sector professionals across all four regions we studied. This extends the picture from our 2023 survey, which found generative AI already in use by 22\% of UK public sector workers around a year after ChatGPT's release \cite{bright2025generative}. What we observed then as an emerging pattern in a single country now appears, some sixteen months later, across every region we studied. Individual use outpaces formal organisational adoption everywhere, most starkly in Europe, where 65\% of respondents use generative AI but only 19\% report that their organisation has adopted AI.

This gap reflects the different paths by which the technology reaches workers and institutions. Consumer-facing tools allow individuals to bypass the procurement processes through which public sector technology has traditionally arrived. Public bodies, meanwhile, remain constrained by the legacy systems, regulatory requirements and caution that have long slowed digital change \cite{dunleavy2006digital, cordella2015egovernment}. Supervisors may occupy an important position here. They report higher generative AI use than non-supervisors (77\% vs 62\%), greater optimism about it (73\% vs 60\%) and greater awareness of relevant guidelines. This suggests they may act as an informal bridge between individual practice and institutional policy, and that they are a natural channel through which organisations could reach their staff.

Perhaps surprisingly, frontline workers are slightly more likely than non-frontline workers to use generative AI (71\% vs 64\%), even though frontline roles are often assumed to face greater barriers to technology adoption. Our earlier study argued that generative AI's appeal lies partly in the personal agency it offers: workers can tailor it to their own needs rather than adapting to a system built for a single purpose \cite{bright2025generative}. This flexibility may lower the threshold for adoption across a wider range of roles. Our task-level data, however, shows that frontline use remains concentrated on administration (emails, documents, reports and meeting minutes), while tasks involving direct work with the public are the least common use in every profession. Frontline workers are also less aware of AI guidelines than their non-frontline colleagues. This could be a concern given they have a direct interface with the public. The professions with the lowest uptake include Social Welfare and Housing (49\%) and Public Safety and Security (53\%). These were also the sectors that had the lowest uptake in our 2023 survey: social care (11\%) and the emergency services (12\%). 

South Africa is a distinctive case. It reports the highest levels of adoption, sustained use, optimism and guideline awareness of any region, and it is the only region where around half of respondents say their organisation has formally adopted AI. Part of this lead can be attributed to sample composition: the South African sample is younger, includes more supervisors and has more respondents in Technology and Digital Services, all of which are associated with higher adoption. Broader factors such as national digital strategy and labour market conditions may also play a role. However, more than half of South African users say generative AI has increased what is expected of them at work, and more than 70\% say it has changed the scope of their day-to-day work, far more than in any other region. 

This points to a broader tension across our sample. Around 80\% of generative AI users report a positive effect on their productivity, a level very similar to that in 2023, and two-thirds (67\%) say it frees up time for more creative or enjoyable work. Yet nearly a third of users, also report that it has raised what is expected of them. In 2023, NHS workers in our survey believed that well-used AI could save them the equivalent of a full day each week, and we suggested this might strengthen the case for a four-day working week \cite{bright2025generative}. Our new findings suggest a less comfortable possibility: that time saved is absorbed by higher demands rather than returned to workers. 

Concern about job loss, meanwhile, remains secondary for most respondents. In 2023 only 16\% feared AI would replace their job, and we find that concern about job security is now only slightly higher among non-users than users (23\% vs 16\%). Underlying much of this is a persistent problem of risk and governance. More than 70\% of respondents are concerned about the security and privacy of AI systems, more than half see bias as a serious concern in their field, and 45\% of users worry about the accuracy of generative AI outputs.

These concerns coexist with a widespread sense that guidance is lacking, which more than half of respondents report. Our guideline-awareness results suggest the problem lies less in the absence of policy than in its failure to reach workers. The UK Generative AI Framework for HM Government was published in January 2024, just over a year before our survey, yet only 8\% of UK respondents were aware of it. In 2023, only 32\% of UK users felt their workplace had clear guidance \cite{bright2025generative}, and the intervening frameworks do not appear to have changed workers' experience much. The low rate of organisational adoption compounds the problem: where employers have not provided sanctioned tools, workers are likely to be using consumer tools under consumer terms, which is precisely where privacy and security concerns are most pressing. Publishing frameworks, in other words, is not enough if they do not reach the frontline workers who are already using these tools.

 Users are twice as likely as non-users to be optimistic about AI in their field (75\% vs under 40\%). This is consistent with the idea that direct experience shapes attitudes more strongly than public debate, but our cross-sectional data cannot rule out the reverse: that those who are already optimistic are the ones who take the technology up. These findings carry several implications for policy and practice:

First, rather than relying on publication alone, governments and employers should disseminate guidance through supervisors, embed it in induction and training, and attach it to the tools workers actually use. Second, frontline workers use generative AI at least as much as others but know less about the rules governing it, and professions such as Social Welfare and Housing combine lower use with the highest concern about bias. Targeted training and support in these areas would address risk where it is greatest. Third, if productivity gains are being absorbed by rising expectations, measures of output alone will overstate the benefits. Employers should track workload and wellbeing alongside efficiency.

Our 2023 paper closed by raising the question of who will provide these technologies in future, and whether governments should rely on a few large technology companies or invest in their own models \cite{bright2025generative}. That question has become more pressing. ChatGPT, Copilot and Gemini now dominate public sector use, largely through individual rather than institutional adoption. 

These findings should be read alongside the study's limitations. Our data is self-reported, descriptive and cross-sectional, so we cannot establish causal relationships. Our sample, recruited via Prolific and unweighted, may over-represent workers comfortable with digital tools. Self-reported awareness of named guidelines is likely to be inflated (Section~3.6).
\paragraph*{Data Availability Statement.} 
The code and datasets are available upon request and will be published publicly at a later date. 

\bibliographystyle{apalike}
\bibliography{main}

\newpage
\section{Appendix}
\label{sec:appendix}



\begin{table}[htbp]
\centering
\small
\begin{tabular}{llrr}
\toprule
\textbf{Variable} & \textbf{Category} & \textbf{n} & \textbf{Freq} \\
\midrule

\multirow{3}{*}{Gender}
& Female & 2127 & 63\% \\
& Male & 1191 & 35\% \\
& others & 44 & 2\% \\
\addlinespace
\hline \addlinespace

\multirow{6}{*}{Age}
& Under 30 & 980 & 29\% \\
& 30--40 & 1175 & 35\% \\
& 41--50 & 658 & 20\% \\
& 51--60 & 420 & 12\% \\
& Above 60 & 125 & 4\% \\
& Prefer not to say & 4 & 0\% \\
\addlinespace
\hline \addlinespace

\multirow{9}{*}{Country}
& United Kingdom & 999 & 30\% \\
& United States & 950 & 28\% \\
& South Africa & 845 & 25\% \\
\cmidrule{2-4}
& \textbf{All European countries} & \textbf{568} & \textbf{17\%} \\
& Germany & 91 & 3\% \\
& Portugal & 89 & 3\% \\
& Spain & 89 & 3\% \\
& Italy & 88 & 3\% \\
& Poland & 80 & 2\% \\
& Ireland & 72 & 2\% \\
& France & 59 & 2\% \\
\addlinespace
\hline \addlinespace

\multirow{11}{*}{Profession}
& Education & 1132 & 34\% \\
& Healthcare and Public Health & 682 & 20\% \\
& Public Safety  and Security & 326 & 10\% \\
& Legislation and Governance & 265 & 8\% \\
& Technology and Digital Services & 191 & 6\% \\
& Social Welfare and Housing & 188 & 6\% \\
& Infrastructure and Transportation & 154 & 5\% \\
& Economic and Financial Management & 152 & 5\% \\
& Environment and Natural Resources & 118 & 4\% \\
& Other & 154 & 5\% \\
\addlinespace
\hline \addlinespace


\multirow{2}{*}{Worker type}
& Frontline role & 2125 & 63\% \\
& Non-frontline role & 1237 & 37\% \\
\addlinespace
\hline \addlinespace

\multirow{2}{*}{Supervisor}
& Non-supervisor & 2007 & 60\% \\
& Supervisor & 1355 & 40\% \\
\addlinespace

\bottomrule
\end{tabular}
\caption{Sample characteristics}
\label{tab:sample_characteristics}
\end{table}


\begin{table}[htbp]
\centering
\small
\begin{tabular}{llrrrr}
\toprule
\textbf{Variable} & \textbf{Value} & 
\makecell{\textbf{Generative} \\ \textbf{AI}} & 
\makecell{\textbf{Perceptive} \\ \textbf{AI}} & 
\makecell{\textbf{Decision}  \\ \textbf{Support}} & 
\makecell {\textbf{Overall} \\ \textbf{AI}}  \\

\midrule

\multirow{3}{*}{Gender}
& Male & 70\% & 36\% & 39\% & 77\% \\
& Female & 67\% & 34\% & 35\% & 73\% \\
& Non-binary & 71\% & 24\% & 26\% & 76\% \\
\midrule
\addlinespace

\multirow{5}{*}{Age}
& Under 30 & 77\% & 47\% & 48\% & 84\% \\
& 30--40 & 68\% & 33\% & 36\% & 74\% \\
& 41--50 & 65\% & 23\% & 25\% & 68\% \\
& 51--60 & 59\% & 25\% & 28\% & 66\% \\
& Above 60 & 56\% & 30\% & 32\% & 66\% \\
\midrule
\addlinespace

\multirow{10}{*}{Profession}
& Technology and Digital Services & 88\% & 63\% & 65\% & 93\% \\
& Education & 79\% & 31\% & 30\% & 83\% \\
& Infrastructure and Transportation & 69\% & 40\% & 39\% & 77\% \\
& Economic and Financial Management & 62\% & 30\% & 47\% & 73\% \\
& Healthcare and Public Health & 64\% & 41\% & 49\% & 72\% \\
& Environment and Natural Resources & 64\% & 30\% & 26\% & 71\% \\
& Legislation and Governance & 59\% & 29\% & 30\% & 66\% \\
& Public Safety and Security & 53\% & 34\% & 31\% & 63\% \\
& Social Welfare and Housing & 49\% & 15\% & 19\% & 53\% \\
& Other & 63\% & 30\% & 34\% & 68\% \\
\midrule
\addlinespace

\multirow{2}{*}{Frontline Worker}
& Frontline role & 71\% & 37\% & 39\% & 77\% \\
& Non-frontline role & 64\% & 30\% & 32\% & 71\% \\
\midrule
\addlinespace

\multirow{2}{*}{Supervisor Worker}
& Supervisor & 77\% & 45\% & 48\% & 84\% \\
& Non-supervisor & 62\% & 27\% & 29\% & 68\% \\
\midrule
\addlinespace

\multirow{4}{*}{Region}
& South Africa & 84\% & 67\% &  62\% & 91\% \\
& United States & 70\% & 29\% & 32\% & 74\% \\
& Europe & 65\% & 20\% &  26\% & 74\% \\
& United Kingdom & 55\% & 19\% & 25\% & 61\% \\

\bottomrule
\end{tabular}
\caption{AI use by demographics for different AI types}
\label{tab:ai_use_characteristics}
\end{table}








\begin{figure}
    \centering
    \includegraphics[width=1\linewidth]{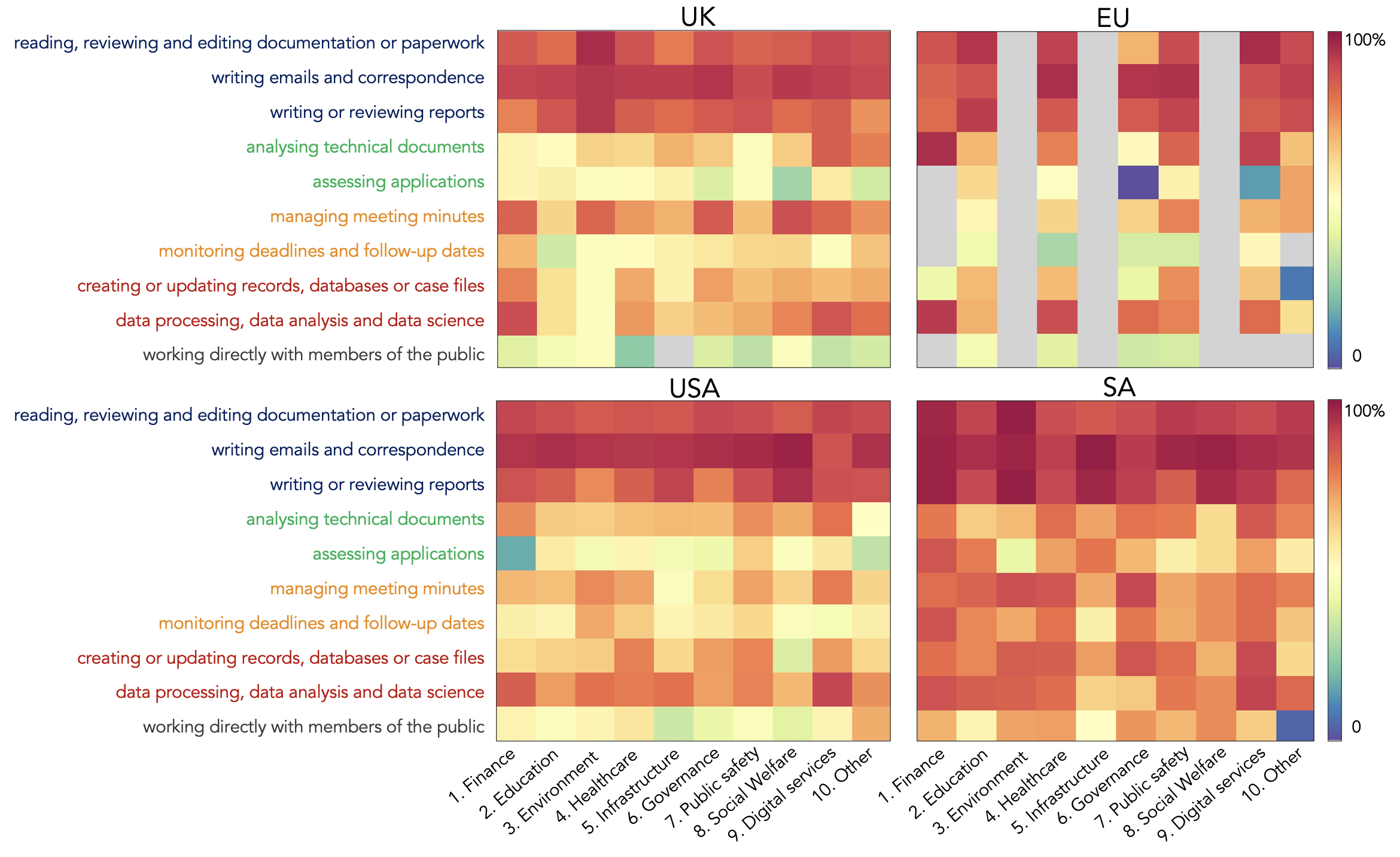}
    \caption{AI use by profession, region and type of tasks}
    \label{fig:use_cases}
\end{figure}

\begin{figure}
    \centering
    \includegraphics[width=1.1\linewidth]{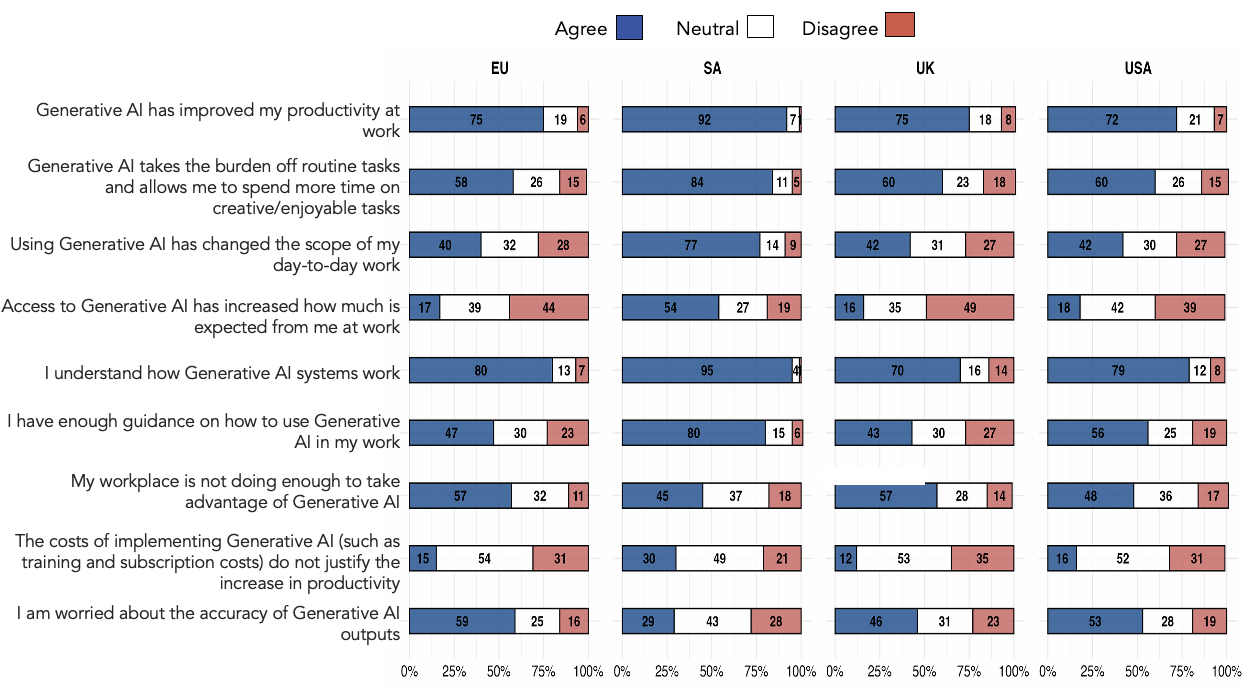}
     \caption{Opinion of AI users by region}
    \label{fig:AI_user_region_opinion}
\end{figure}
\begin{figure}
    \centering
    \includegraphics[width=1.1\linewidth]{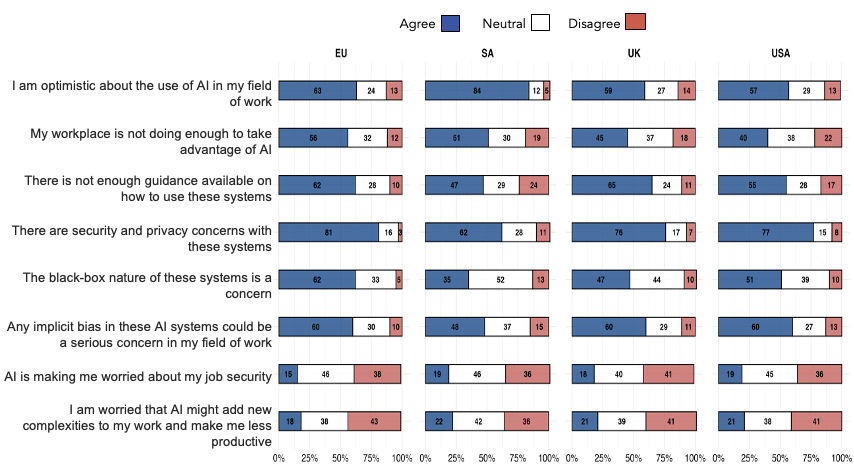}
    \caption{Opinions of all respondents by region}
    \label{fig:all_opinion_country}
\end{figure}

\begin{figure}
    \centering
    \includegraphics[width=1.1\linewidth]{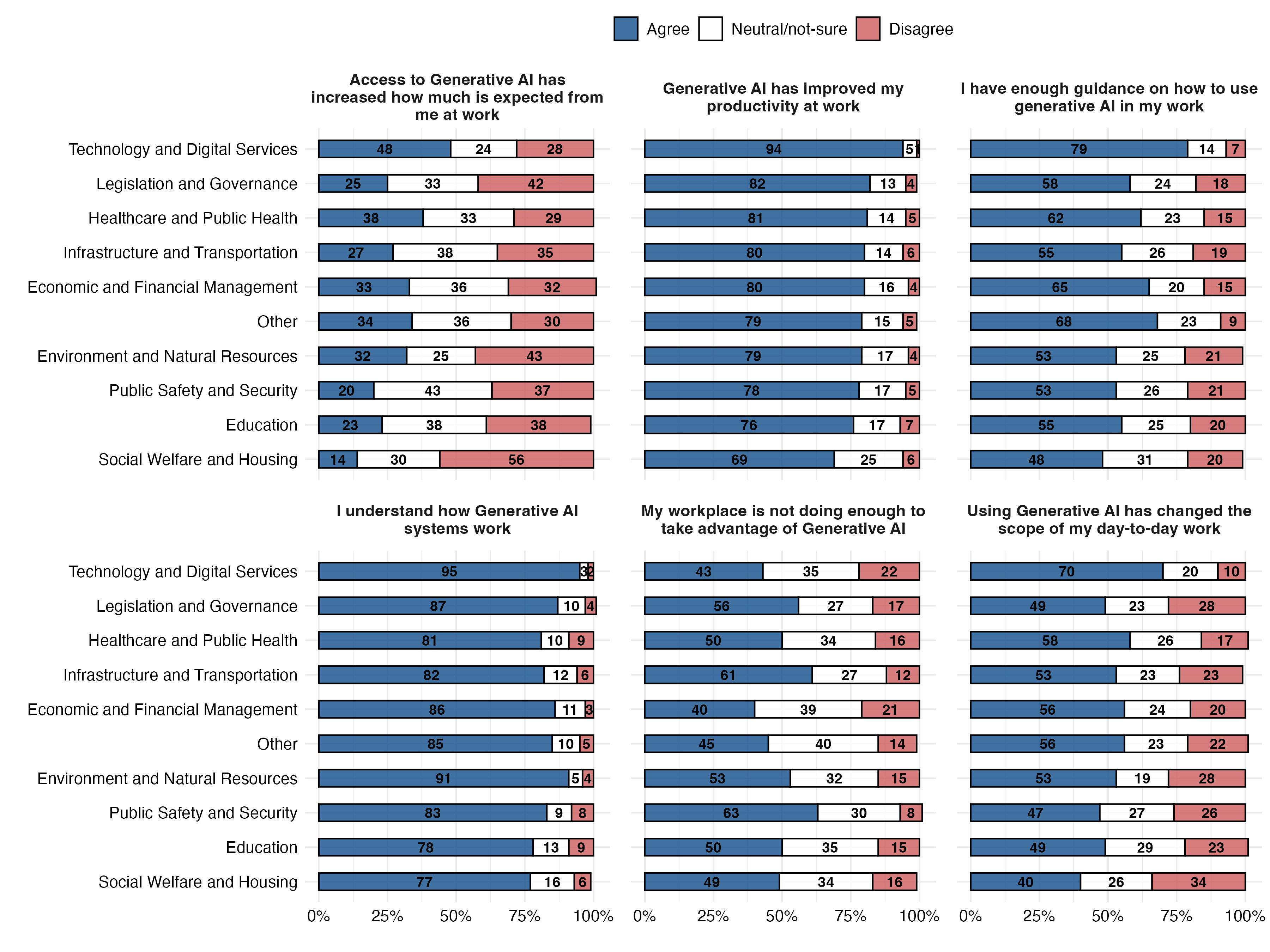}
    \caption{Opinions of AI users by profession}
    \label{fig:ai_opinion_prof}
\end{figure}
\begin{figure}
    \centering
    \includegraphics[width=1.1\linewidth]{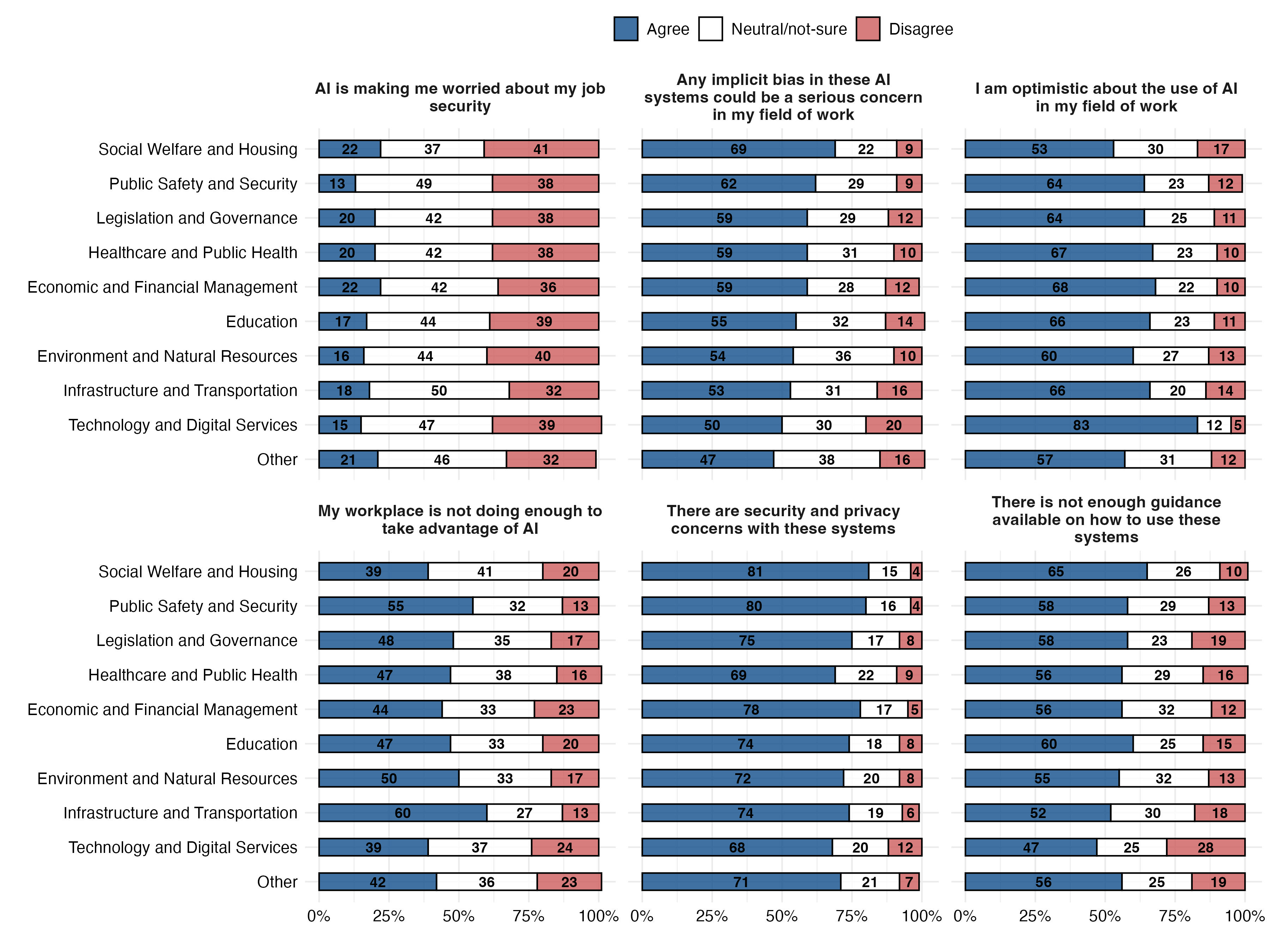}
    \caption{Opinions of all respondents by profession}
    \label{fig:all_opinion_prof}
\end{figure}
\begin{figure}
    \centering
    \includegraphics[width=1.1\linewidth]{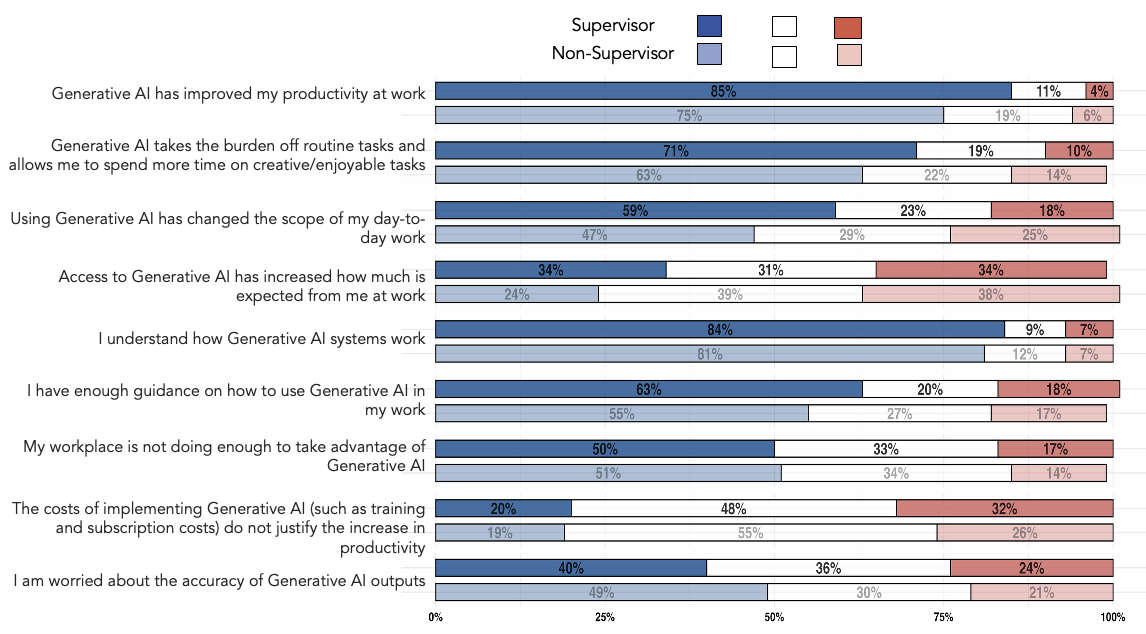}
     \caption{Opinion of AI users by role}
    \label{fig:AI_user_supervisor_opinion}
\end{figure}

\begin{figure}
    \centering
    \includegraphics[width=1\linewidth]{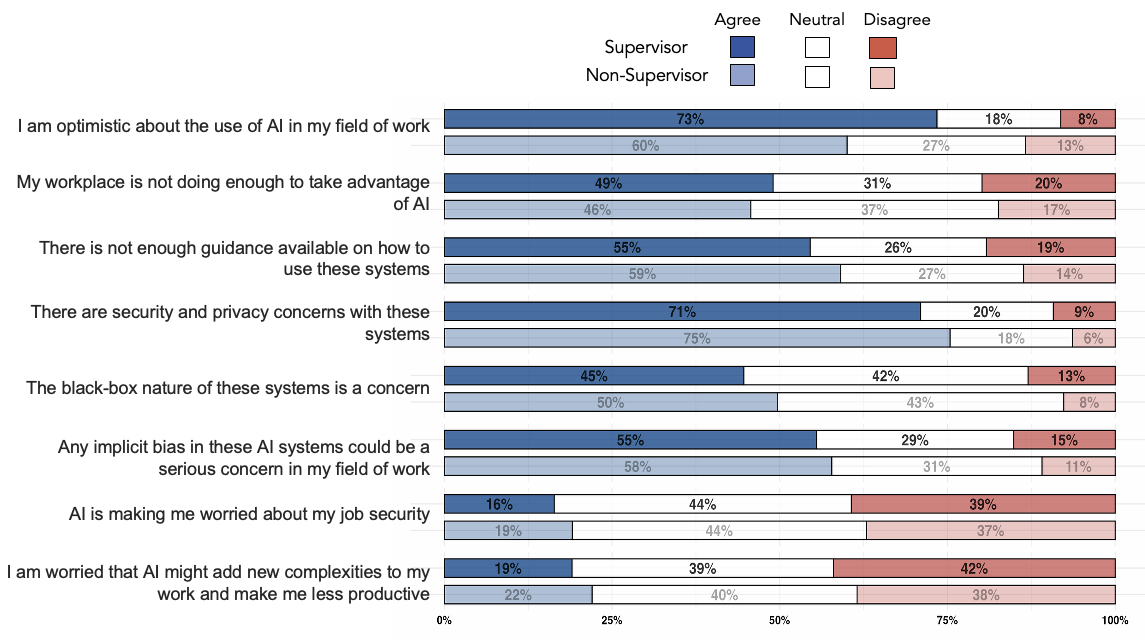}
    \caption{Opinions  of all respondents by role}
    \label{fig:all_opinion_supervisor}
\end{figure}

\end{document}